\documentclass[aps,showpacs,superscriptadress,preprint]{revtex4}
\usepackage{graphicx}
\usepackage{amsmath}

\begin{document}
\title{Nuclear matter and neutron matter for improved quark mass
       density- dependent model with $\rho$ mesons}

\author{
Chen Wu$^1$ and Ru-Keng Su$^{1,2,3}$\footnote{rksu@fudan.ac.cn}}
 \affiliation{
\small 1. Department of Physics, Fudan University,Shanghai 200433, P.R. China\\
\small 2. CCAST(World Laboratory), P.O.Box 8730, Beijing 100080, P.R. China\\
\small 3. Center of Theoretical Nuclear Physics,\\
National Laboratory of Heavy Ion Collisions, Lanzhou 730000, P.R.
China}

\begin{abstract}
A new improved quark mass density- dependent model including u, d
quarks, $\sigma$ mesons,  $\omega$ mesons and $\rho$ mesons is
presented. Employing this model, the properties of nuclear matter,
neutron matter and neutron star are studied. We find that it can
describe above properties successfully. The results given by the new
improved quark mass density- dependent model and by the quark meson
coupling model are compared.
\end{abstract}

 \maketitle

\section{Introduction}
In our previous papers [1-6], a new quark meson coupling model bases
on quark mass density- dependent(QMDD) model is presented. The QMDD
model suggested  by Fowler, Raha and Weiner [7] firstly assuming
that the masses of u, d and s quarks(and the corresponding
antiquarks) satisfy:
\begin{eqnarray}
m_{q} = \frac{B}{3n_{B}}(i = u,d,\bar u,\bar d)   \label{mq1}
\end{eqnarray}
\begin{eqnarray}
m_{s,\bar s } = m_{s0}+\frac{B}{3n_{B}}     \label{mq2}
\end{eqnarray}
where $n_{B}$ is the baryon number density, $m_{s0}$ is the current
mass of the strange quark, and  $B$ is the  bag constant. As was
explained in Refs.[1, 2, 5, 8], the ansatz Eqs. (1) and (2)
corresponds to a quark confinement hypothesis because when
$V\rightarrow \infty, n_B \rightarrow 0$ and $m_q \rightarrow
\infty$, it prevents the quark goes to infinity. It is shown that
the properties of strange quark matter in the QMDD model are nearly
the same as those obtained in the MIT bag model [9, 10]. But the
basic difference is that instead of the MIT bag boundary condition,
we have the density- dependent masses of quarks in QMDD model
according to Eqs. (1) and (2). It means that the ansatz Eqs. (1) and
(2) can replace MIT bag boundary condition and get the nearly the
same results.

  Quark- meson  coupling(QMC) model suggested by Guichon [11]
firstly is a famous hybrid quark meson model which can describe many
physical properties of nuclear matter and nuclei successfully [12].
In this model, the nuclear system was suggested as a collection of
MIT bag and mesons. The interactions between quarks and mesons are
limited within the MIT bag regions. As was pointed in Refs. [1, 2,
6], this model has two major shortcomings: (1) It is  a permanent
quark confinement model because the MIT bag boundary condition
cannot be destroyed by temperature and density. Therefore, it cannot
describe the quark deconfinement phase transition. (2) It is
difficult to do nuclear many-body calculation beyond mean field
approximation(MFA) by means of QMC model, because we cannot find the
free propagators of quarks and mesons  easily. The reason is that
the interactions between quarks and mesons are limited within the
bag regions, the multireflection of quarks and mesons by MIT bag
boundary must be taken into account for getting the free
propagators. These two shortcomings come from MIT bag constrain all.

To overcome these two shortcomings, we suggested an improved quark
mass density- dependent(QMDD) model in Refs. [1-6]. We added the
$\sigma$- meson and $\omega$- meson to improve the QMDD model.
Instead of the MIT bag, after introducing the nonlinear interaction
of $\sigma$-mesons and qq$\sigma$ coupling, we construct a
Friedberg- Lee soliton bag in nuclear system. The quark masses are
still density- dependent. The interactions between quarks and mesons
are extended to the whole system. Since the MIT bag constraint is
given up, our improved  QMDD(IQMDD) model can describe the quark
deconfinement phase transition [6] and do the nuclear many- body
calculations beyond MFA in principle. We have proved that our model
can successfully describe the saturation properties, the equation of
state, the compressibility and the effective nucleon mass of
symmetric nuclear matter and give a reasonable critical temperature
of quark deconfinement [1-6].

The motivation of this paper is to extend our study to asymmetric
nuclear matter, especially to the neutron matter and the neutron
star. It means that we must consider the isosipn dependence and
distinguish the u-quark and d-quark. We will add the isospin vector
$\rho$ mesons to  improve the  IQMDD model in this paper. We hope to
compare the results of IQMDD model with those obtained by QMC model
and QHD-II model for neutron matter and neutron star. In order to
find their differences and similarity explicitly, we will use the
same approximation as that of the QMC model [13] in our
calculations. Though the study of neutron star employing QMC model
in Ref. [14] is too simple, but in order to exhibit the basic
differences between the IQMDD model and the QMC model, we still
consider the neutron star by using the same approximation as Ref.
[14].

The organization of this paper is as follows. In the next section,
we give the main formulas of the IQMDD model under the mean field
approximation. The main formulas of neutron stars in also included.
In the third section, some numerical results are presented. The last
section contains a summary and discussions.

\section{Formulas of the IQMDD model with $\rho$ meson}
 \subsection {IQMDD model for nuclear matter}
 The Lagrangian density of IQMDD model with $\sigma, \omega, $ and $\rho$ mesons is :
 \begin{eqnarray}
\mathcal{L} =
\bar{\psi}[i\gamma^{\mu}\partial_\mu-m_{q}+g^q_\sigma\sigma-g^q_\omega\gamma^\mu\omega_\mu
-g^q_\rho \gamma^\mu \vec{\tau}\cdot \vec{\rho}^\mu
]\psi+\frac{1}{2}\partial_{\mu}\sigma\partial^{\mu}\sigma \hskip
0.0in \nonumber\\-U(\sigma)
-\frac{1}{4}F_{\mu\nu}F^{\mu\nu}+\frac{1}{2}m_{\omega}^2\omega_\mu\omega^\mu
+\frac{1}{2}m_{\rho}^2\vec{\rho_\mu} \cdot \vec{\rho^\mu}
-\frac{1}{4}\vec{G_{\mu\nu}}\vec{G^{\mu\nu}}
\end{eqnarray}
where
\begin{eqnarray}
 U(\sigma) = \frac{1}{2}m_\sigma^2\sigma^2 +
\frac{1}{3}b\sigma^3+ \frac{1}{4}c\sigma^4+ B,
\end{eqnarray}
\begin{eqnarray}
 -B = \frac{m_\sigma^2}{2}
\sigma_v^2+\frac{b}{3}\sigma_v^3+\frac{c}{4}\sigma_v^4,
\end{eqnarray}
\begin{eqnarray}
\sigma_v=\frac{-b}{2c}[1+\sqrt{1-4m_\sigma^2 c/b^2}],
\end{eqnarray}
and the quark mass $m_q$(q = u, d) is given by Eq. (1).
 $m_\sigma$
and $m_\omega$ are the masses of $\sigma$ and $\omega$ mesons,
$F_{\mu\nu}=\partial_\mu \omega_\nu-
\partial_\nu \omega_\mu$, $\overrightarrow{G}_{\mu\nu}=\partial_\mu \overrightarrow{\rho}_\nu-
\partial_\nu \overrightarrow{\rho}_\mu$, $g_\sigma^q$, $g_\omega^q$ and $g_\rho^q$ are the
coupling constants  between quark and $\sigma$ meson, quark and
$\omega$ meson and quark and $\rho$ meson respectively.

The equation of motion for quark field  under MFA in the whole space
is
\begin{eqnarray}
[i \gamma \cdot \partial - (m_q- g_\sigma^q \bar\sigma) - \gamma^0
(g_\omega^q \bar\omega+ \frac{1}{2}g_\rho^q \tau_z \bar\rho)]\psi=0
\end{eqnarray}
where $\bar\sigma, \bar\omega$ are the mean field values of the
$\sigma$ field and the corresponding time component  of $\omega$
field respectively, $\bar\rho$ is the mean field value of the time
component in the third direction of isospin for $\rho$ field,
$\tau_z$ is the third component of the Pauli matrix. The effective
quark mass $m_q^{*}$ is given by:
\begin{eqnarray}
 m_q^{*}=m_q-g_\sigma^q\bar{\sigma}
\end{eqnarray}
In nuclear matter, three quarks constitute a Freidberg-Lee soliton
bag [15], and the effective nucleon mass is obtained from the bag
energy and reads:
\begin{eqnarray}
 M_N^* = \Sigma_q E_q =\Sigma_q \frac{4}{3}\pi R^3
\frac{\gamma_q}{(2\pi)^3}\int_0^{K_F^q}\sqrt{{m^*_q}^2+k^2}
(\frac{dN_q}{dk})dk
\end{eqnarray}
where $\gamma_q$ is the quark degeneracy, $K_F^q$ is Fermi energy of
quarks. $dN_q/dk$ is the density of states for various quarks in a
spherical cavity. The expression of $dN_q/dk$ adopted in this paper
con be found in Ref. [16].

The Fermi energy $K_F^q$  of quarks is given by
\begin{eqnarray} 3= \frac{4}{3}\pi R^3 n_B
\end{eqnarray}
where $n_B$ satisfies
\begin{eqnarray} n_B =\Sigma_q \frac{ \gamma_q}{(2\pi)^3} \int_0^{K_F^q}(\frac{dN_q}{dk}) dk
\end{eqnarray}
The bag radius $R$ is determined by the equilibrium condition for
the nucleon bag:
\begin{eqnarray} \frac{\delta M^*_N}{\delta R}=0
\end{eqnarray}

In nuclear matter, the total energy density  and pressure density
read
\begin{eqnarray}
\varepsilon_{matter}&=& \frac{\gamma_N}{(2\pi)^3} (\int_0^{K_F^p}+
\int_0^{K_F^n}) \sqrt{{M_N^*}^2+p^2} dp^3 +
\frac{g_\omega^2}{2m_\omega^2}\rho_B^2 \nonumber\\ &&
+\frac{1}{2}m_\sigma^2 \bar{\sigma}^2 +\frac{1}{3}b \bar{\sigma}^3 +
\frac{1}{4}c \bar{\sigma}^4+ \frac{g_\rho^2}{8m_\rho^2 }{\rho_3^2}
\end{eqnarray}
and
\begin{eqnarray}
p_{matter}&=& \frac{1}{3}\frac{\gamma_N}{(2\pi)^3} (\int_0^{K_F^p}+
\int_0^{K_F^n}) \frac{p^2}{\sqrt{{M_N^*}^2+p^2}} dp^3 +
\frac{g_\omega^2}{2m_\omega^2}\rho_B^2 \nonumber\\ &&
-\frac{1}{2}m_\sigma^2 \bar{\sigma}^2 -\frac{1}{3}b \bar{\sigma}^3 -
\frac{1}{4}c \bar{\sigma}^4+ \frac{g_\rho^2}{8m_\rho^2 }{\rho_3^2}
\end{eqnarray}
where $\gamma_N=2$ is degeneracy of proton or neutron, $K_F^p$ and
$K_F^n$ is Fermi momenta of proton and neutron, and $\rho_3$ is the
difference between the proton and neutron densities, respectively.
Therefore
\begin{eqnarray}
 \rho_p =\frac{1}{3\pi^2} {K_F^p}^3, \rho_n =\frac{1}{3\pi^2}
 {K_F^n}^3,
\end{eqnarray}
Where $\rho_p$ and $\rho_n$ is the density of proton and neutron
respectively, and   the density of nuclear matter $\rho_B$ resds
\begin{eqnarray}
 \rho_B =\rho_p+\rho_n
\end{eqnarray}

In Eqs. (13, 14), $g_\omega$ and $g_\rho$ are  the coupling
constants between nucleon and  $\omega$ meson, nucleon and $\rho$
meson respectively. They satisfies $g_\omega=3g_\omega^q$ and
$g_\rho = g_\rho^q$ [13]. In MFA, the $\bar\omega$ is determied by
baryon number conservation
\begin{eqnarray}
 \bar\omega = \frac{g_\omega\rho_B}{m_\omega^2}
\end{eqnarray}

 As that of the QMC model [13],  $\bar\sigma$ and $\bar\rho$
 are given by the thermodynamics conditions:
\begin{eqnarray}
(\frac{\partial \varepsilon_{matter}}{\partial \bar\sigma})_{R,
\rho_B} = 0, \ \ \ \  and  \ \ \ \ (\frac{\partial
\varepsilon_{matter}}{\partial \bar\rho})_{R, \rho_B} = 0
\end{eqnarray}
respectively.  Therefore,  $\bar\rho$  is expressed by
\begin{eqnarray}
\bar\rho =\frac{g_\rho}{2m_\rho^2} \rho_3
\end{eqnarray}
and  $\bar\sigma$  is given by
\begin{eqnarray}
m_\sigma^2 \bar\sigma+b\bar\sigma^2+c\bar\sigma^3 = -
\frac{\gamma_N}{(2\pi)^3} (\int_0^{K_F^p}+\int_0^{K_F^n})
\frac{M_N^*}{\sqrt{{M_N^*}^2+p^2}} d^3 p (\frac{\partial
M_N^*}{\partial \bar\sigma})_R
\end{eqnarray}
 Eqs. (13-20) form a complete set of equations and we can solve
numerically. Our numerical results will be shown in the next
section.

Noting that the left hand side of Eq. (20) is a cubic order function
of $\bar\sigma$ , except one solution $\bar\sigma=0$, there are
still two solutions. This is a general character for adding a
nonlinear scalar $\sigma$ field and using MFA to a physical model
[17]. But as was pointed in Ref. [1], we can prove that one of these
solutions corresponds to unstable and unphysical branch, and the
other corresponds a stable soliton solution. Hereafter we give up
the unphysical solution and consider the physical solution only.

We  note that the expression for the total energy density, Eq. (13),
is very similar to that of QHD-II model and QMC model. The
differences comes from the effective nucleon mass Eqs. (9) and (1),
and the self-consistency condition for $\sigma$ field, Eq. (20). Let
us consider the  self-consistency condition and $(\frac{\partial
M_N^*}{\partial \bar\sigma})_R$ further. Using the same argument as
that of Ref. [13], we find that the
 $(\frac{\partial M_N^*}{\partial \bar\sigma})_R$ can be
expressed as
 \begin{eqnarray}
(\frac{\partial M_N^*}{\partial \bar\sigma})_R = -g_\sigma \times
\left( \begin{array}{c} 1\\ C_1(\bar\sigma)\\ C_2(\bar\sigma)
\end{array}\right)
\,\,\,\, for \left( \begin{array}{c} QHD-II\\ QMC\\ IQMDD
\end{array}\right) model
\end{eqnarray}
where the expression of scalar density factor $C_1(\bar\sigma)$ for
QMC model can be found in Ref. [13]. For IQMDD model,
$C_2(\bar\sigma)$ can be obtained numerically. The curves of
$C_1(\bar\sigma)$ and $C_2(\bar\sigma)$ will be shown in Sec. 3.

\subsection {IQMDD model for neutron star}
We now turn to  investigate the neutron  matter and neutron star for
the IQMDD model. Since the aim of this paper is to compare  the
IQMDD model and the QMC model, we use the same approximation to
study the neutron star as that of the QMC model [14]. More detail
treatment of neutron star such as the   phase transition for the
quark matter and the neutron matter, the contribution of hyperon and
 etc, has been neglected. Two basic assumptions: the neutron star matter
is  charge neutrality and reaches to the $\beta$-equilibrium, are
adopted [14]. Since we assume that  the nucleons and light leptons
exist in the neutron star only, charge neutrality is expressed as
\begin{eqnarray}
\rho_p = \sum_{l=e, \mu}\rho_l,
\end{eqnarray}
where $\rho_i$ is the number density of particle $i(= p, e, \mu)$.
Under $\beta$-equilibrium, the processes
\begin{eqnarray}
n \rightarrow p + e^- + \bar\nu_e, p + e^- \rightarrow n  + \nu_e
\end{eqnarray}
occur at the same rate. This condition can be satisfied when the
chemical potentials before and after the decay are same. The
chemical potential of each particle reads
\begin{eqnarray}
\mu_n &=& \sqrt{{K_F^n}^2+ {m_N^*}^2}+ g_\omega \bar\omega-
\frac{1}{2}g_\rho \bar\rho \\
\mu_p &=& \sqrt{{K_F^p}^2+ {m_N^*}^2}+ g_\omega \bar\omega+
\frac{1}{2}g_\rho \bar\rho \\
\mu_l &=& \sqrt{K_l^2 + m_l^2}
\end{eqnarray}
where $K_l$ is the Fermi momentum of the lepton $l(e, \mu)$. The
chemical equilibrium condition is expressed as
\begin{eqnarray}
\mu_n &=& \mu_p + \mu_e, \\ \mu_e &=& \mu_\mu
\end{eqnarray}
Once the solution has been found, the equation of state(EoS) can be
calculated from
\begin{eqnarray}
\varepsilon_{} &=& \frac{\gamma_N}{(2\pi)^3} (\int_0^{K_F^p}+
\int_0^{K_F^n}) \sqrt{{M_N^*}^2+p^2} dp^3 +
\frac{g_\omega^2}{2m_\omega^2}\rho_B^2+\frac{1}{2}m_\sigma^2
\bar{\sigma}^2 \nonumber\\&&+\frac{1}{3}b \bar{\sigma}^3 +
\frac{1}{4}c \bar{\sigma}^4 + \frac{g_\rho^2}{8m_\rho^2 }{\rho_3}^2
+ \frac{1}{\pi^2}\sum_{l} \int_0^{k_l} \sqrt{k^2 + m_l^2} k^2 dk,
\end{eqnarray}
\begin{eqnarray}
p_{} &=& \frac{1}{3}\frac{\gamma_N}{(2\pi)^3} (\int_0^{K_F^p}+
\int_0^{K_F^n}) \frac{p^2}{\sqrt{{M_N^*}^2+p^2} }dp^3 +
\frac{g_\omega^2}{2m_\omega^2}\rho_B^2-\frac{1}{2}m_\sigma^2
\bar{\sigma}^2 \nonumber\\&& -\frac{1}{3}b \bar{\sigma}^3 -
\frac{1}{4}c \bar{\sigma}^4 + \frac{g_\rho^2}{8m_\rho^2 }{\rho_3}^2
+ \frac{1}{3\pi^2}\sum_{l} \int_0^{k_l}  \frac{k^4}{\sqrt{k^2 +
m_l^2}} dk,
\end{eqnarray}

Using the Oppenheimer and Volkoff equation
\begin{eqnarray}
\frac{d p(r) }{d r} &=&
-\frac{Gm(r)\varepsilon}{r^2}(1+\frac{p}{\varepsilon C^2})(1+
\frac{4\pi r^3 p}{m(r)C^2}) (1-\frac{2Gm(r)}{rC^2})^{-1}
\end{eqnarray}
\begin{eqnarray}
dM(r) = 4\pi r^2 \varepsilon(r) dr
\end{eqnarray}
where G is gravitational constant and C is the velocity of light,
and the equation of state for neutron matter given by Eqs. (29),
(30) and (9), we can study the physical behavior of neutron star for
IQMDD model.

\section{numerical result}
Before numerical calculation, let us discuss the parameters in IQMDD
model. First, we choose $m_\omega=783$ MeV, $m_\rho=770$ MeV and
$m_\sigma=509$ MeV as that of Ref. [18]. Fixing the nucleon mass
$M_N=939$ MeV, we get $B=174$ MeV fm$^{-3}$. Obviously, the
behaviors at the saturation point must be explained  for a
successful model. It reveals that nuclear matter saturates at a
density $\rho_0=0.15$ fm$^{-3}$  with a binding energy per particle
$E/A= -15$ MeV at zero temperature, and the compression constant to
be about $K(\rho_0)=210$ MeV. Therefore we fixed $ g_\omega^q=2.44,
g^q_\sigma=4.67, b=-1460$ MeV to explain above data. In addition,
the symmetry energy coefficient $a_{sym}$ satisfies
\begin{eqnarray}
a_{sym} = \frac{1}{2}(\frac{\partial^2(\varepsilon/\rho)}{\partial
\alpha^2})_{\alpha=0} = (\frac{g_\rho}{m_\rho})^2
 \frac{k_0^3}{12\pi^2} + \frac{k_0^2}{6 \sqrt{k_0^2+ {M^\star_N}^2}}
\end{eqnarray}
 where
 \begin{eqnarray}
 \alpha = \frac{\rho_n-\rho_p}{\rho_n+\rho_p}, k_0=1.42fm^{-1}
\end{eqnarray}

Using the data $a_{sys}$=33.2 MeV we fix $g_\rho=9.07$. The
parameters used to calculate and the results of $K$ and $M_N^*$ for
IQMDD model are shown in Table 1. For comparison, we also show the
corresponding parameters for QMC model in Table 1. The data and
results for QMC model  are taken from Ref. [13]. We see in Table 1
all parameters and results are very similar for these two models.
Their differences are not remarkable.
\\

\begin{tabular}
{p{1.7cm}p{2.0cm}p{2.0cm}p{2.0cm}p{2.0cm}p{2.0cm}p{2.0cm}}
\multicolumn{6}{c}{TABLE 1.  Comparison of calculated quantities in IQMDD and QMC model  } \\
\hline\hline  & ${g_\sigma^q}$ &${g_\omega^q}$ &${g_\rho^q}$ &R(fm)
 &K(MeV)   &$M^*_N(\rho_0)(MeV)$
\\\hline
QMC &5.53  &1.26  &8.44 &0.80  &200 &851   \\
IQMDD &4.67  &2.44 &9.07 &0.85  &210 &775 \\
 \hline\hline
\end{tabular}
\\

Our results for symmetric nuclear matter and neutron matter are
shown in Fig. 1-3. The scalar density factor $C(\bar\sigma)$ as a
function of $\bar\sigma$ is shown in Fig. 1  where the dashed curve
refers to $C_1(\bar\sigma)$  and solid curve to $C_2(\bar\sigma)$
 respectively. This factor plays an essential role
to demonstrate the main character of quark structure for different
models. We see from Fig. 1 that the scalar density factors
$C_1(\bar\sigma)$ and $C_2(\bar\sigma)$ are both  smaller than
unity(QHD-II model) and decrease when $\bar\sigma$ increases. In
particular, $C_2(\bar\sigma)$ is located between the line of unity
and the curve of $C_1(\bar\sigma)$. It means that the values of main
physical quantities given by IQMDD model will almost located between
the values given by QHD-II model and by QMC model. Our results
confirm this conclusion.

In Fig. 2, we draw the curves of energy per baryon vs. baryon number
density for both symmetric nuclear matter and for neutron matter
respectively. We see that ignoring $\rho$ meson coupling yields a
smaller bound state around $\rho_B \sim 0.10 fm^{-3} \sim
0.66\rho_0$ in the dotted curve, but it becomes unbound solid line
when the $\rho$ meson contribution is introduced. The saturation
curve for symmetric nuclear matter is shown by dash-dotted curve in
Fig. 2. In fact, the behavior of curves in Fig. 2 is very similar to
that of QMC model, but the corresponding value of $\rho_B$ is
0.60$\rho_0$ for QMC model and 0.66$\rho_0$ for IQMDD model. (See
Fig. 2 of Ref. [13])

The equation of state for neutron matter is shown in Fig. 3 where
the dashed curve presents the result when the $\rho$ meson
contribution is ignored and the solid curve corresponds to the full
calculation. We see the contribution of $\rho$ meson is important
for the EoS. After comparing with the results of QHD-II model and
QMC model, we come to a conclusion that the shape of equation of
state for neutron matter  in IQMDD model is qualitatively similar to
that of QHD-II model and QMC model, it is softer than that of QHD-II
model but harder than that of the QMC model, as is indicated in Fig.
1 by the behavior of scalar density factor.

 Having shown the IQMDD model can provide an successful description
for nuclear and neutron matter, we would like to study the structure
and composition of neutron stars for IQMDD model. We will show that
it can successfully describe the neutron star.

The EoS is given by Eqs. (9), (29) and (30) for IQMDD model when the
neutron star matter reaches to $\beta$ equilibrium. The curve of EoS
is shown in Fig. 4. In Fig. 5, we show the particle population
including $n, p, e, \mu$ for different density by solid($n$),
dashed($p$), dotted($e$), dash-dotted($\mu$) curves respectively.
The mass of neutron star in units of sun mass $M/M_{\bigodot}$ as a
function of central density $\varepsilon_c$ is plotted in  Fig. 6.
In Fig. 7 we show the mass radius relation of the neutron star. The
maximum mass of neutron star $M_{max}$ found in Fig. 7 is 1.73
$M_{\bigodot}$. It is smaller than the value of 2.2$M_{\bigodot}$
given by QMC model [14]. We would like to emphasize that the above
treatment for neutron star is too rough. Therefore, the value of
$M_{max}$ is not important. Our aim is to demonstrate that all
curves shown in Fig. 4-7 are in agreement with those given by QMC
model qualitatively [14]. We come to a conclusion that perhaps the
IQMDD model is a good candidate to replace the QMC model.
\section{Summary and discussion}
In summary, we have added the $\rho$ meson to the IQMDD model to
study the asymmetric nuclear matter, especially, the neutron matter
and neutron star. The $u, d,$ quarks, nonlinear scalar $\sigma$
meson field, $\omega$ meson field, $\rho$ meson field and the
corresponding quark mesons couplings are including in the new IQMDD
model. The isospin effect has been considered by introducing
isovector $\rho$ mesons in this model. After fixing the parameters
by the experimental values such as the massed of nucleon, $\sigma$
meson, $\omega$ meson, $\rho$ meson; saturation point, compression
constant and the symmetry energy, under MFA, we have investigate the
physical properties of nuclear matter and neutron matter. We found
that the results given by IQMDD model are similar to that of QMC
model. The values of the main physical quantities for neutron matter
and nuclear matter given by IQMDD model locate in the regions
between the values given by the QHD-II model and by the QMC model.
Employing the IQMDD model, we have studied the neutron star and also
found its properties almost agree with these given by QMC model. We
conclude that the new IQMDD model with $\rho$ meson is successful
for describing the nuclear matter and neutron matter. Perhaps it can
play the role to replace the QMC model.

\begin{acknowledgments}
The author C.W.  wish to thank Prof. Ren-Xin Xu discussions and
correspondence. This work is supported in part by the National
Natural Science Foundation of People's Republic of China.
\end{acknowledgments}

\begin{figure}[tbp]
\includegraphics[width=14cm,height=20cm]{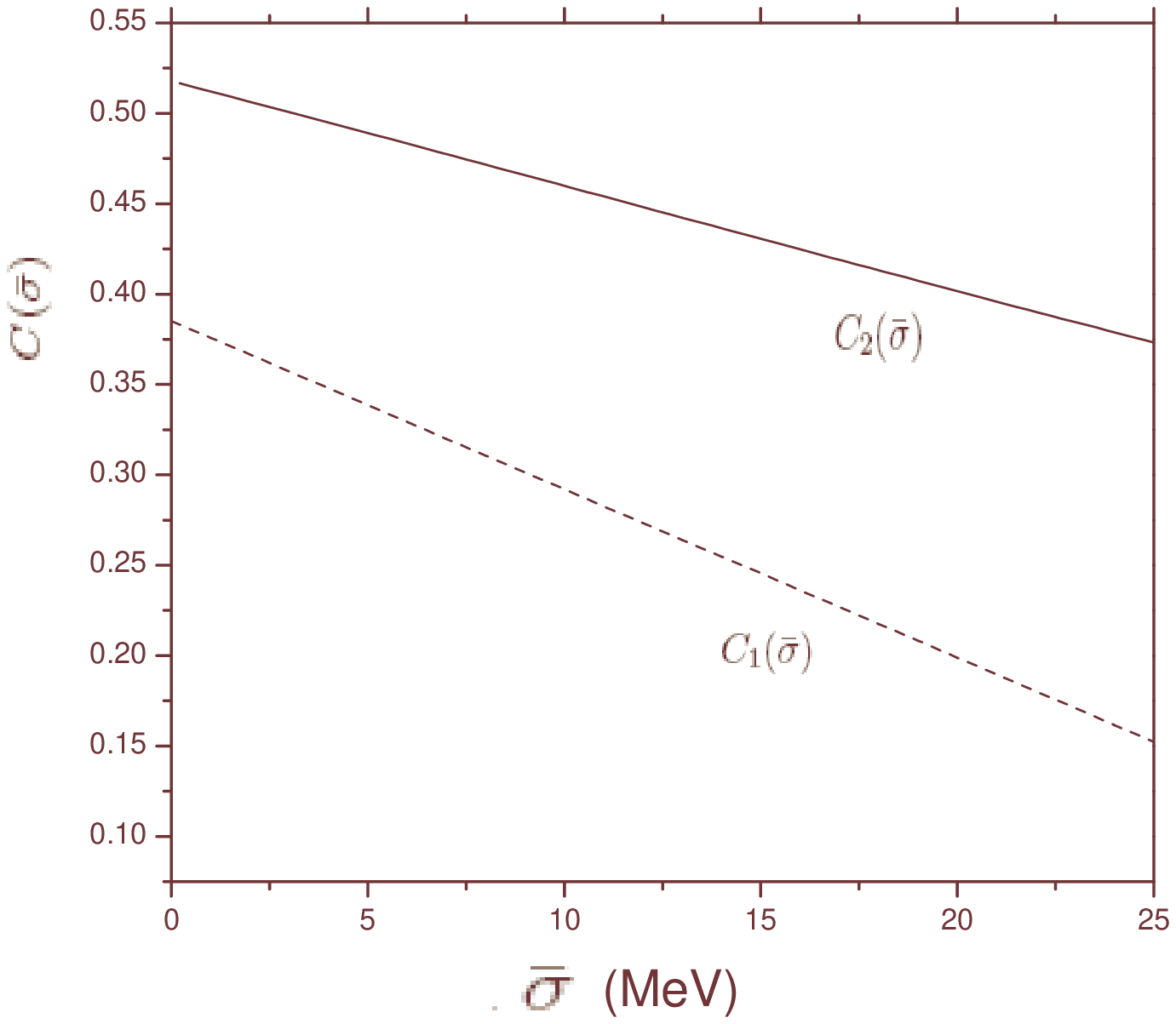}
\caption{Scalar density factors $C_2(\bar\sigma)$ and
$C_1(\bar\sigma)$ as a function of $\bar\sigma$ for IQMDD model and
QMC model respectively.}

\end{figure}

\begin{figure}[tbp]
\includegraphics[width=14cm,height=20cm]{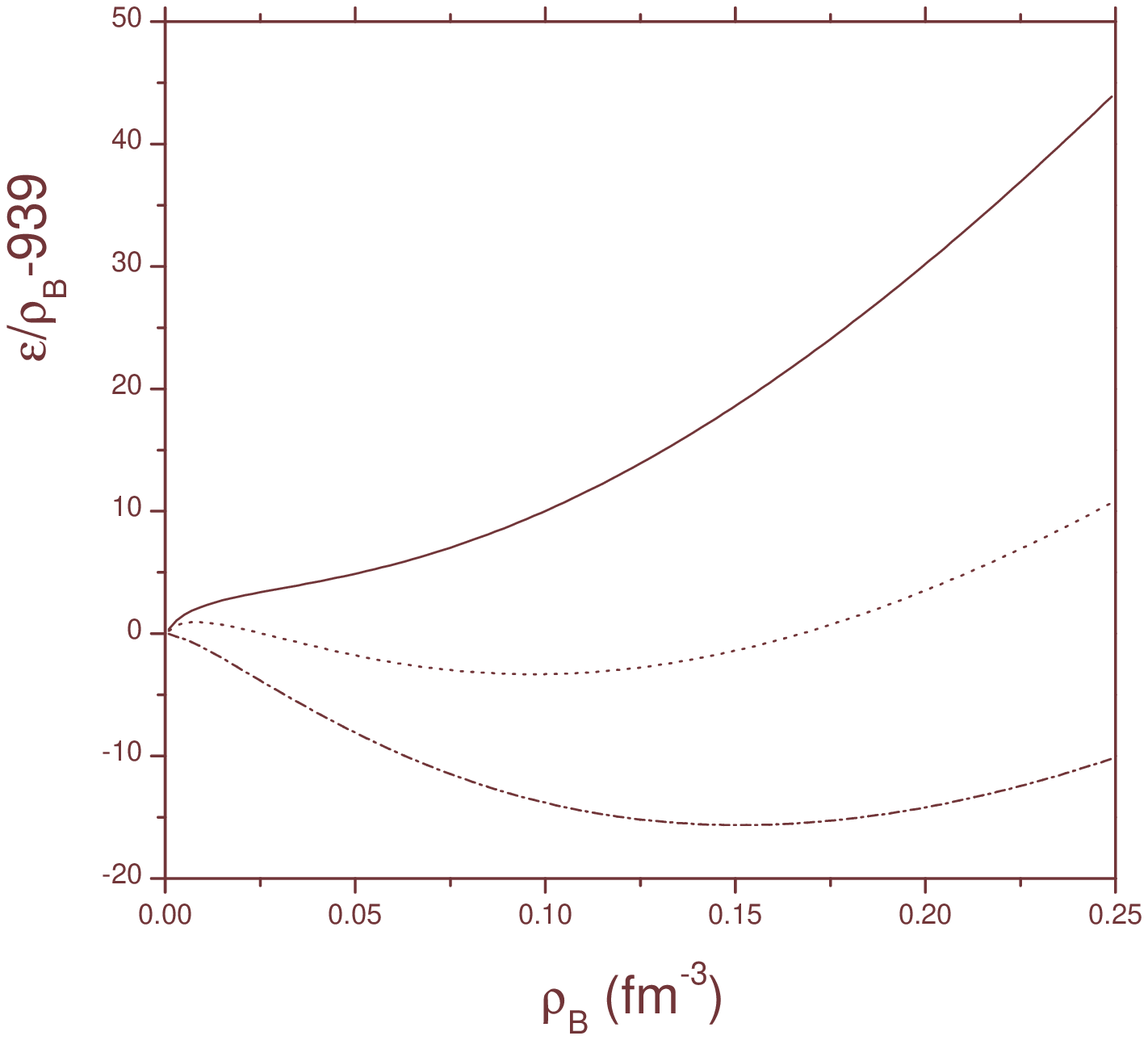}
\caption{Energy per nucleon for symmetric nuclear matter and for
neutron matter. The  dash-dotted curve is the saturation curve for
nuclear matter. The solid curve(with $\rho$ meson) and the dotted
curve(omit $\rho$ meson) show the results for neutron matter.}
\end{figure}

\begin{figure}[tbp]
\includegraphics[width=14cm,height=20cm]{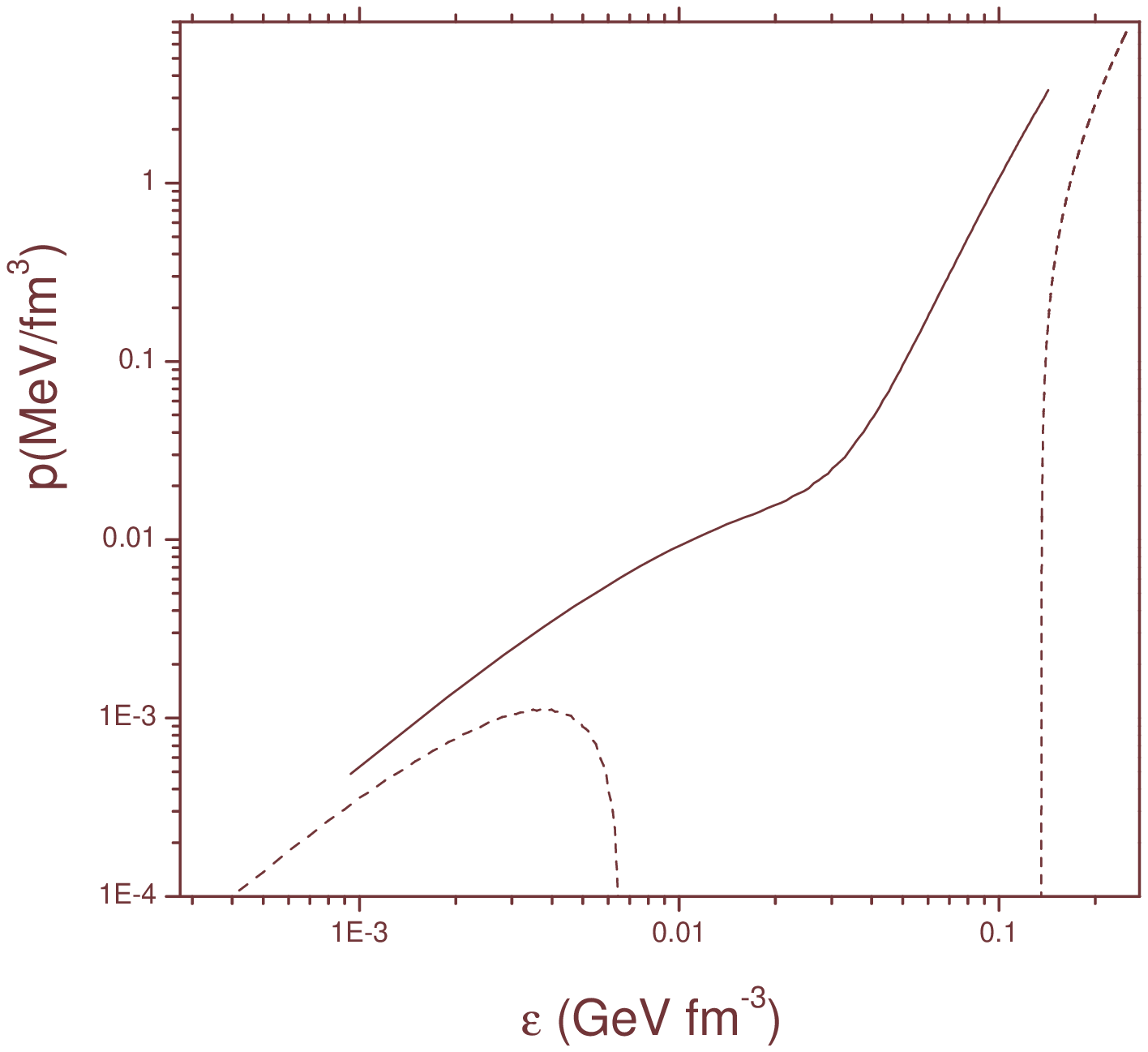}
\caption{The equation of state for neutron matter. The dashed line
show the result for omitting $\rho$ meson, while the solid line
corresponds to full consideration.}
\end{figure}

\begin{figure}[tbp]
\includegraphics[width=14cm,height=20cm]{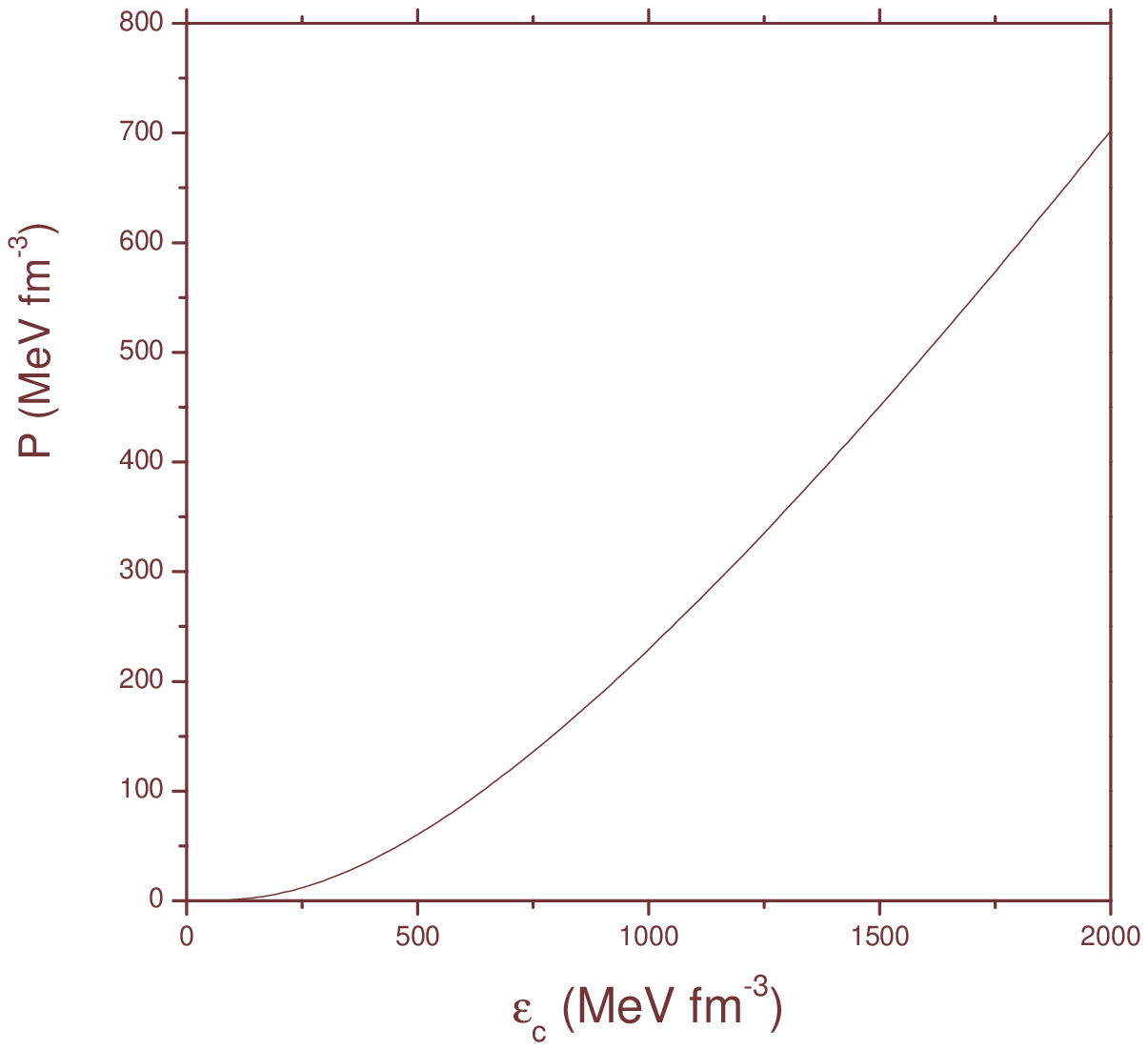}
\caption{Equation of state for $\beta$- equilibrium neutron star
matter. }
\end{figure}

\begin{figure}[tbp]
\includegraphics[width=14cm,height=20cm]{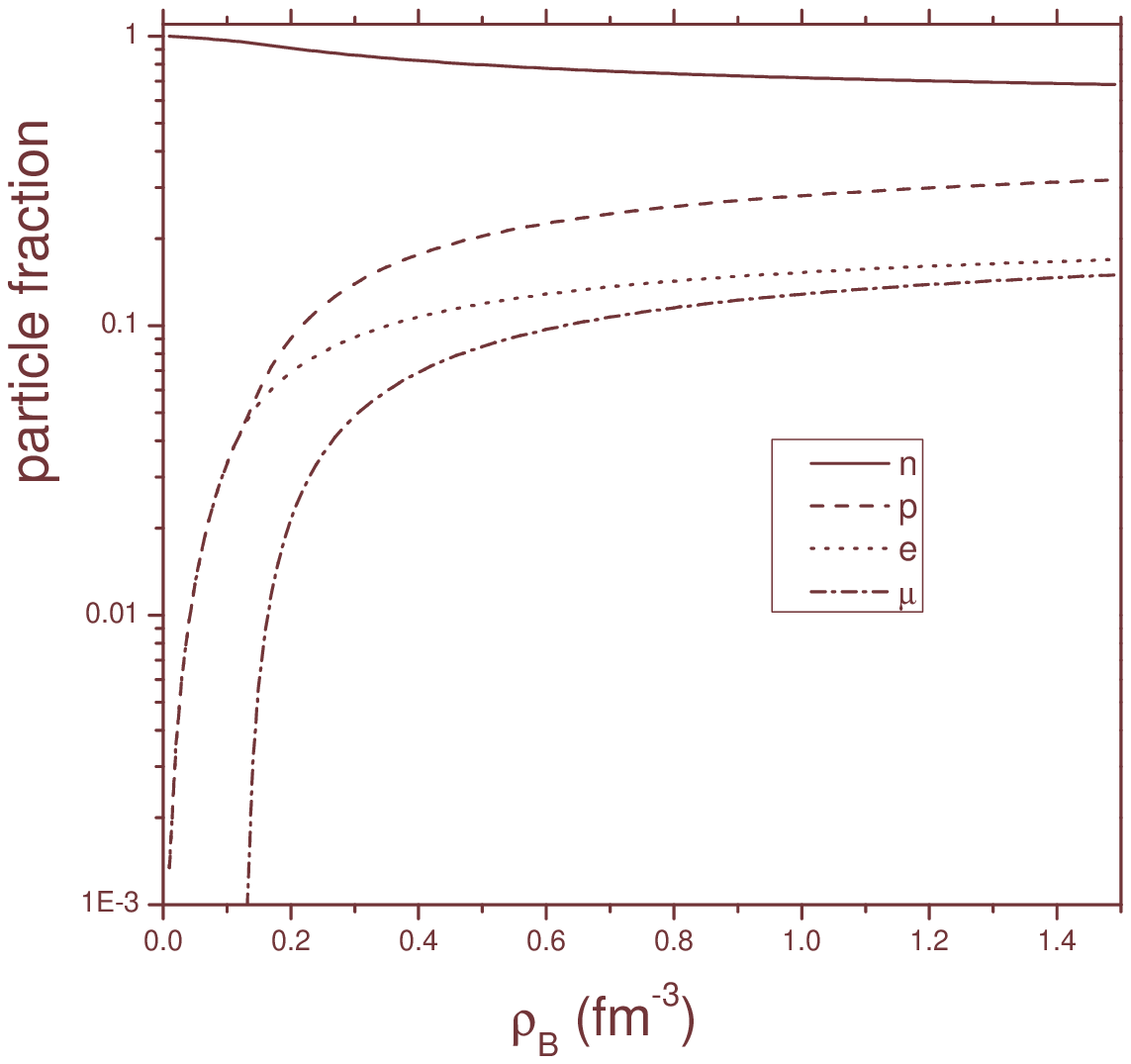}
\caption{Populations in neutron star matter as a function of
density.}
\end{figure}

\begin{figure}[tbp]
\includegraphics[width=14cm,height=20cm]{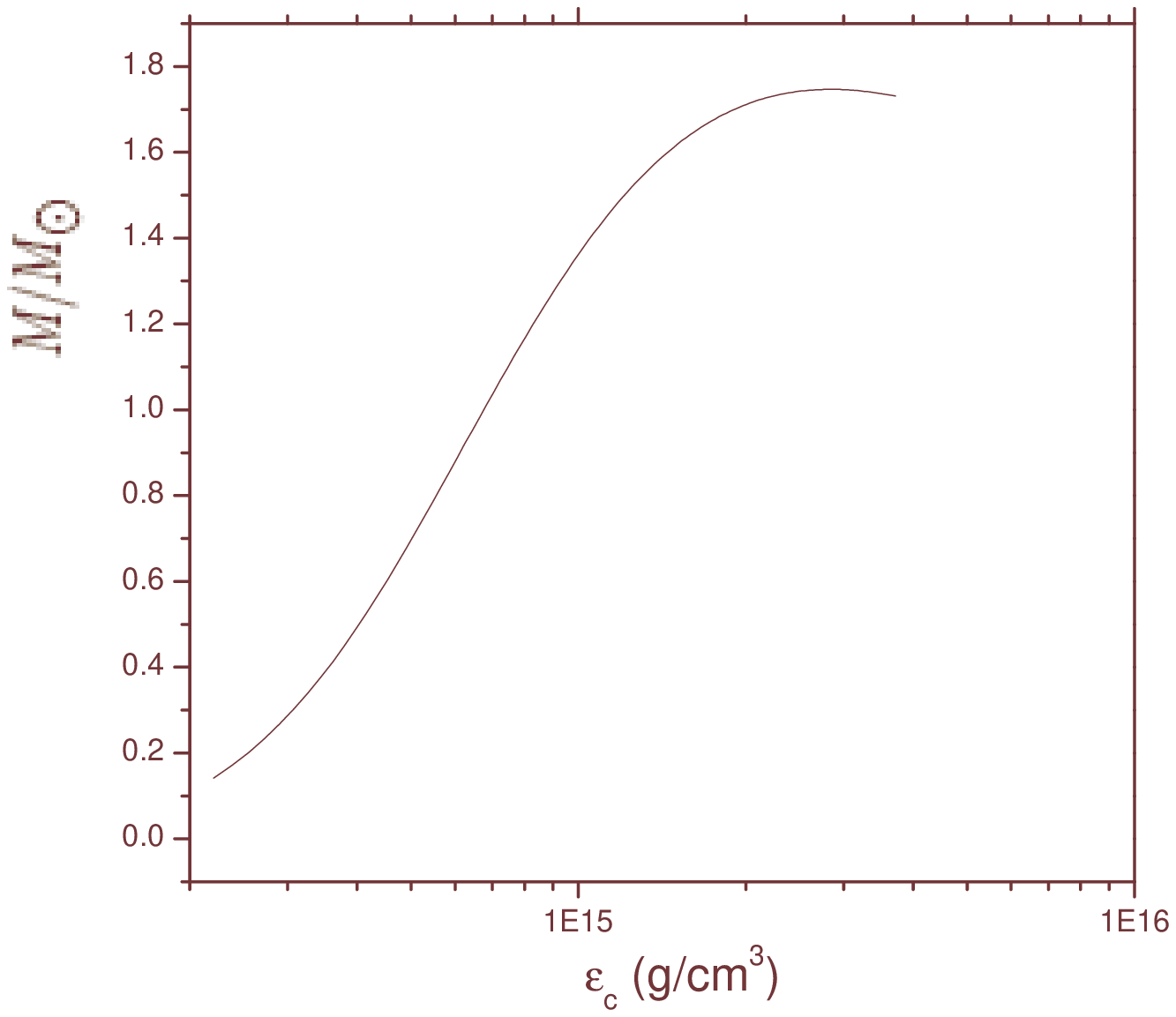}
\caption{Neutron star mass as a function of the central density. }
\end{figure}

\begin{figure}[tbp]
\includegraphics[width=14cm,height=20cm]{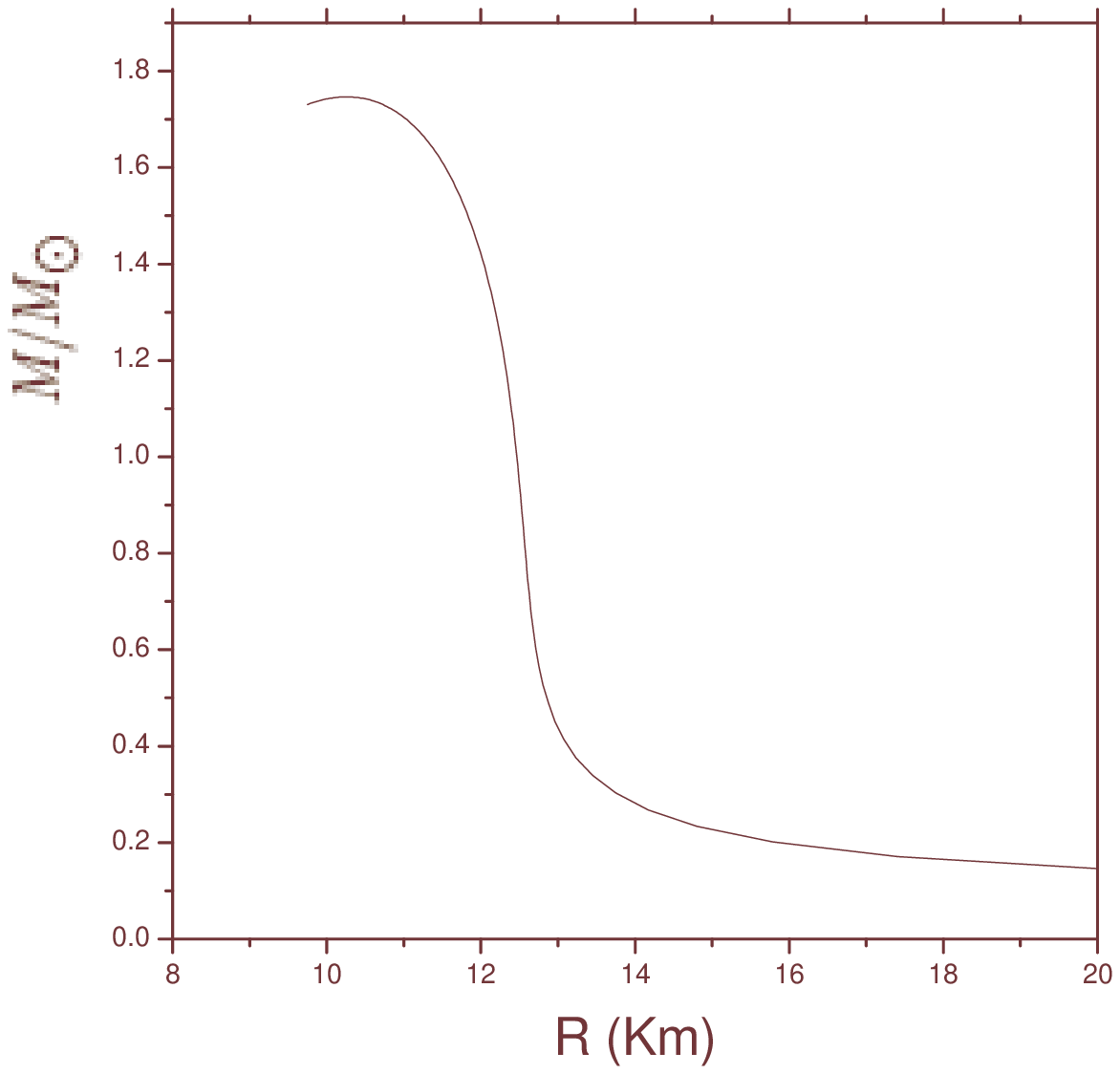}
\caption{Masses of neutron stars vs. their radii. }
\end{figure}

\begin{references}
\bibitem{1} C. Wu and R.K. Su, Phys. Rev. C \textbf{77}, 015203 (2008).
\bibitem{2} C. Wu, W.L. Qian and R.K. Su, Phys. Rev. C \textbf{72}, 035205 (2005).
\bibitem{3} H. Mao, R.K. Su and W.Q. Zhao, Phys. Rev. C \textbf{74}, 055204 (2006).

\bibitem{4} W.L. Qian and R.K. Su, Int. J. Mod. Phys. A \textbf{20}, 1931
(2005).


\bibitem{5} Y. Zhang and R.K. Su, Phys. Rev. C\textbf{65},
035202 (2002); Phys. Rev. C \textbf{67}, 015202 (2003); J. Phys. G
\textbf{30}, 811 (2004).

\bibitem{6}C. Wu and R.K. Su, J. Phys. G(to be published)


\bibitem{7}G.N. Fowler, S. Raha, and R.M. Weiner, Z. Phys. C \textbf{9}, 271
(1981)

\bibitem{8} S. Chakrabarty, S. Raha and B. Sinha, Phys. Lett. B \textbf{229}, 112 (1989).

\bibitem{9}O.G. Benrenuto and G. Lugones, Phys. Rev. D
\textbf{51},  1989 (1995); G. Lugones and O.G. Benrenuto, Phys. Rev.
D \textbf{52},  1276 (1995).

\bibitem{10} S. Yin and R.K. Su, Phys. Rev. C\textbf{77}, 055204 (2008).



\bibitem{11}P.A.M. Guichon, Phys. Lett. B \textbf{200},  235 (1988).

\bibitem{12} K. Saito, K. Tsushima and A.W. Thomas, Prog. in. part.
Nucl. Phys  \textbf{58}, 1 (2007) and Refs therein.

\bibitem{13}K. Saito and A.W. Thomas, Phys. Lett. B
\textbf{327}, 9 (1994).

\bibitem{14} D.P. Menezes, P.K. Panda and C. Providencia,, Phys. Phys. C \textbf{72},
035802 (2005).

\bibitem{15} T.D. Lee, Particle physics and introduction to field theory (Harwood Academic , New York, 1981).

\bibitem{16} Y. Zhang, W.L. Qian, S.Q. Ying and R.K. Su, J. Phys. G \textbf{27},
2241 (2001).

\bibitem{17} B.M. Waldhauser, J.A. Marahu, H. St\"{o}cker and W. Greiner, Phys. Phys. C
\textbf{38},
1003 (1988).

\bibitem{18} R.J. Furnstahl, B.D. Serot and H.B. Tang, Nucl. Phys. A
\textbf{615},
441 (1997).

\end{references}
\end{document}